\documentclass[epjST,final]{svjour}
\usepackage[latin9]{inputenc}
\usepackage{graphicx,color}
\usepackage{hyperref}
\usepackage{booktabs}
\usepackage[normalem]{ulem}
\usepackage{dsfont}
\usepackage[percent]{overpic}
\usepackage{empheq}
\usepackage[numbers,sort&compress]{natbib}

\usepackage{stackengine}
\stackMath

\newcommand{\be}{\begin{equation}}
\newcommand{\ee}{\end{equation}}

\definecolor{mygreen}{rgb}{0,0.5,0}
\definecolor{myblue}{rgb}{0,0,0.75}
\definecolor{mymagenta}{cmyk}{0,1,0,0.12}

\newcommand{\pd}{{\phantom{\dagger}}}

\begin{document}

\title{Coupled-wire constructions: a Luttinger liquid approach to topology}

\author{Tobias Meng\inst{1}\fnmsep\thanks{\email{tobias.meng@tu-dresden.de}}}
\institute{Institut f\"ur Theoretische Physik and W\"urzburg-Dresden Cluster of Excellence ct.qmat, Technische Universit\"at Dresden, 01062 Dresden, Germany}	
\date{\today}

\abstract{
Coupled-wire constructions use bosonization to analytically tackle the strong interactions underlying  fractional topological states of matter. We give an introduction to this technique, discuss its strengths and weaknesses, and provide an overview of the main achievements of coupled-wire constructions.}

\maketitle

\section{Introduction}\label{Sec:intro}
The integer and fractional quantum Hall effects are prototypes for topological states of matter in the non-interaction and interacting cases, respectively \cite{kdp1980,laughlin_state,girvin1999,wen_book,fradkin_book,altland_book,tong_lectures}. Building on the close connections between these prototypes and other topological states has benefited the understanding of topological physics in general: chiral spin liquids can be understood as spin analogues of quantum Hall states \cite{kalmeyer-87prl2095}, $p+i\,p$-superconductors are their superconducting variants \cite{rg2000}, and two-dimensional topological insulators may be viewed as a pair of quantum Hall layers in opposite magnetic fields glued together to globally respect time-reversal symmetry \cite{km2005a,km2005b,bz2006}. Theoretical models for quantum Hall states can thus be generalized to describe many other topological states.

While the integer quantum Hall effect can be understood in a non-interacting picture, fractional topological states necessarily require strong interactions between their elementary constituents (such as electrons or spins). They hence pose a formidable challenge to theory, and a number of ingenious approaches have been used to tackle fractional quantum Hall states, including the clever guesses and numerical verifications of wave functions by Laughlin \cite{laughlin_state}, Haldane's insightful pseudo-potential method \cite{h1983}, or effective field theories \cite{gmp1986,zhk2989}.

A particularly simple and powerful approach to topological states are coupled-wire constructions. They decompose a higher-dimensional system into a collection of one-dimensional subsystems such as electronic quantum wires (hence the name of the technique) or spin chains. Topological states then arise due to suitable couplings between the one-dimensional subsystems. Unlike other approaches based on higher-dimensional field theories, coupled-wire constructions use the powerful bosonization technique to tackle interactions within and between the one-dimensional subsystems.

In the following Sec.~\ref{sec:anisotropy}, we discuss that the anisotropy inherent to coupled-wire constructions leaves the topological properties of quantum Hall states untouched. Sec.~\ref{sec:cw} details how coupled-wire constructions use bosonization to describe fractional quantum Hall states. The main achievements of coupled-wire constructions are summarized in Sec.~\ref{sec:review}. We illustrate the versatility of this approach in Sec.~\ref{sec:general} by discussing generalizations beyond two-dimensional fermionic states of matter, and argue that they are a powerful tool for the future exploration of higher-dimensional fractional phases. Finally, Sec.~\ref{sec:conclusions} concludes the article.

\section{The quantum Hall effect and its anisotropic limit}\label{sec:anisotropy}
Topological systems have an extraordinary robustness to modifications of their microscopic parameters: as long as the modifications do not close the bulk gap, the system remains in the same topological phase, and exhibits the same topological response functions. This is best exemplified  by the extreme robustness and reproducibility of the integer quantum Hall effect.

\begin{figure}
\centering
\hspace*{1cm}\raisebox{2cm}{(a)}~\includegraphics[height=2.25cm]{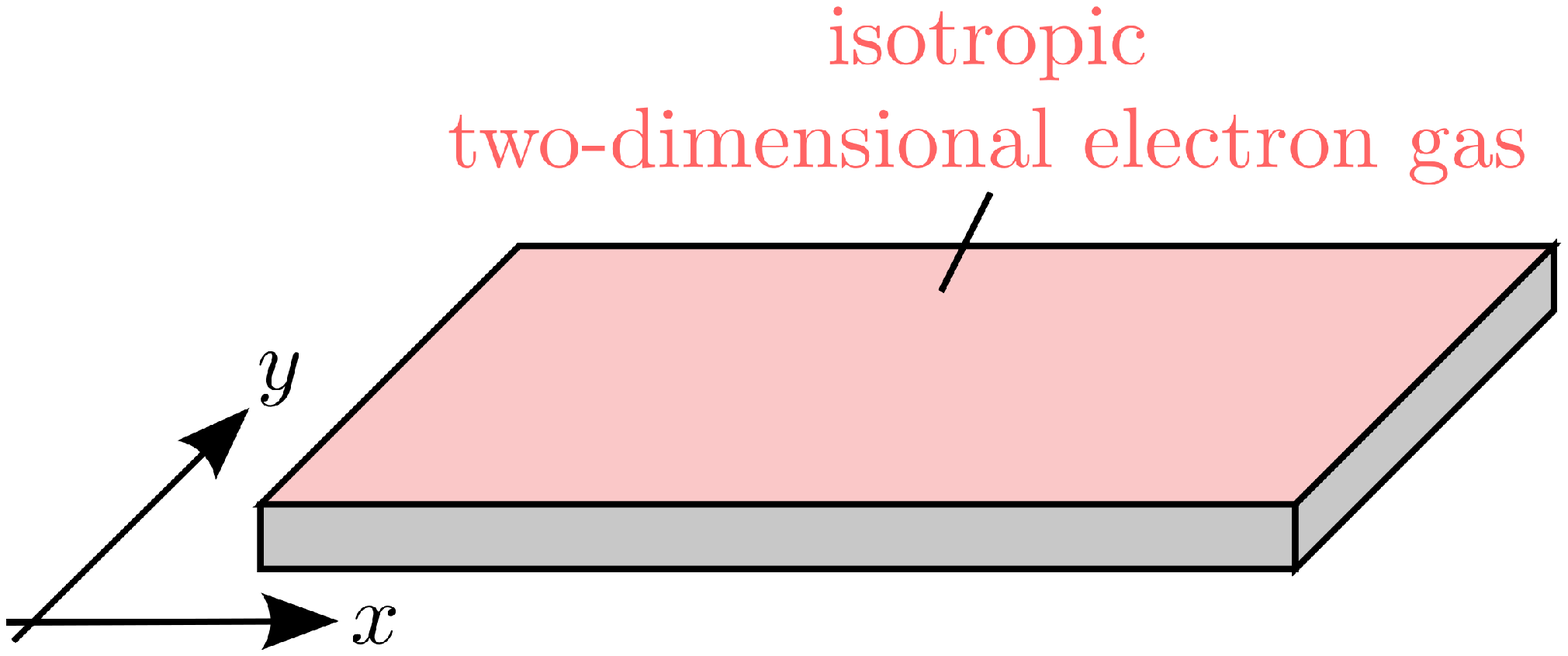}\hspace*{1cm}\raisebox{2cm}{(b)}~
\includegraphics[height=2.25cm]{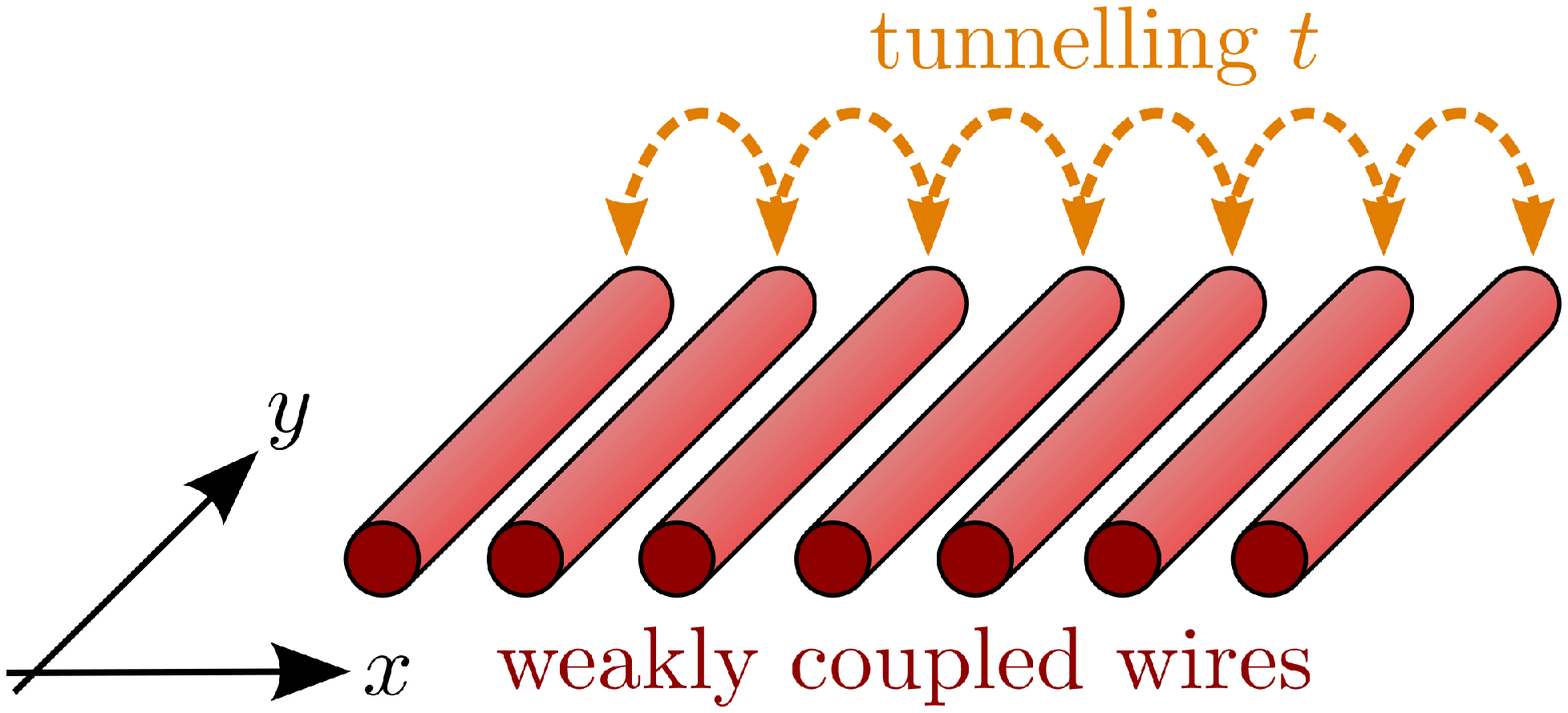}
  \caption{Two-dimensional electron systems at varying anisotropy: panel (a) shows the limit an isotropic two-dimensional electron gas, panel (b) depicts the extremely anisotropic case of an array of weakly tunnel-coupled quantum wires.}
  \label{fig:anisotropy}
\end{figure}

This effect is usually described in terms of  free, spinless electrons in the isotropic $(x,y)$-plane subject to an out-of plane magnetic field $\mathbf{B}=B\,\mathbf{e}_z$, where $\mathbf{e}_i$ is the unit vector in $i$-direction.  Coupled-wire constructions build on the fact that topological states, such as quantum Hall states, are also robust to anisotropy. To illustrate this idea for the integer quantum Hall effect, consider the two-dimensional electron gas shown in Fig.~\ref{fig:anisotropy} (a), and assume that additional electrostatic gates  modulate the system into alternating stripes of high and low electron densities. Because the quantum Hall state can only be destroyed by closing its bulk gap, the Hall effect must remain robust as long as the electrostatic gate potential is much smaller than  the gap. Weak anisotropy is thus irrelevant for the Hall effect. 

\begin{figure}
  \centering
  \raisebox{2.75cm}{(a)}~\includegraphics[height=3cm]{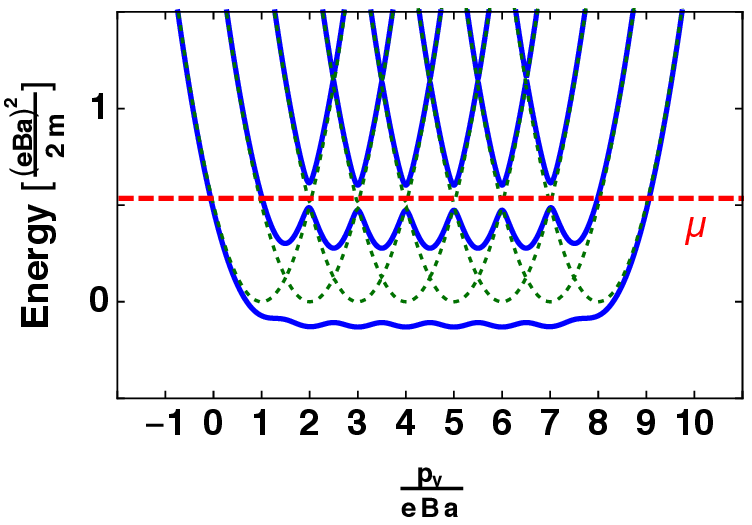}\hspace*{1cm}\raisebox{2.75cm}{(b)}~
  \includegraphics[height=3cm]{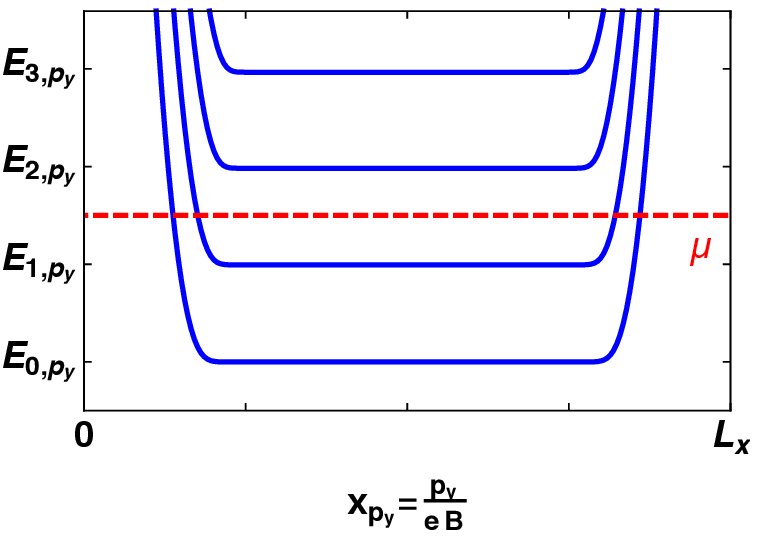}
  \caption{Panel (a) shows the spectrum of an array of eight tunnel-coupled quantum wires extended along $y$ (these systems are sketched in Fig.~\ref{fig:anisotropy}). The green dashed lines depict the parabolic dispersions $\mathcal{E}_j(p_y) = (p_y-e\,B\,j\,a)^2/2m$ with $j=1,\ldots,8$ in the absence of inter-wire tunneling. The solid blue line shows the spectrum for a finite tunnel coupling between neighboring wires. Panel (b) depicts the spectrum of an integer quantum Hall system with finite size $L_x$ along the $x$-direction and periodic boundary conditions along $y$, where $E_{n,p_y}$ denotes the energy of the $n$-th Landau level as a function of $y$-momentum $p_y$. In the Landau gauge $\mathbf{A} = B\,x\,\mathbf{e}_y$, the momentim $p_y$ also labels the $x$-position $x_{p_y} = p_y/eB$ at which a given state is centered. The red dashed line indicates the energy of the chemical potential $\mu$.}
  \label{fig:hall_levels}
\end{figure}

Consider now the ultimate anisotropic limit of weakly coupled quantum wires shown in Fig.~\ref{fig:anisotropy} (b). The wires are extended along $y$, and have an inter-wire distance of $a$. Each wire has a single subband with parabolic dispersion, and is weakly tunnel-coupled to its nearest neighbors. We furthermore apply a magnetic field $\mathbf{B}=B\,\mathbf{e}_z$ and use the Landau gauge $\mathbf{A}(\mathbf{r}) = B\,x\,\mathbf{e}_y$. This system is described by the coupled-wire Hamiltonian
\begin{align}
H_{\text{CW}}&=\sum_{j,p_y}\frac{(p_y-e\,B\,j\,a)^2}{2\widetilde{m}^*}\,c_{j,p_y}^\dagger c_{j,p_y}^\pd + \sum_{j,p_y}\left(t\,c_{j+1,p_y}^\dagger c_{j,p_y}^\pd+\rm{h.c.}\right),\label{eq:hamcwnonint}
\end{align}
where $\widetilde{m}^*$ is the effective electronic mass in the wires, $x=j\,a$ is the $x$-coordinate of the $j$-th wire  (with integer $j$), $t$ denotes the inter-wire tunneling, and $c_{j,p_y}^\dagger$ creates an electron with momentum $p_y$ in wire $j$. The spectrum of $H_{\text{CW}}$ is depicted in Fig.~\ref{fig:hall_levels} (a) for an array of eight wires. At vanishing tunneling, $t=0$, the spectrum consists of one parabola per wire. The dispersion of the $j$-th wire is shifted to $p_y=j\,e \,B\,a$ because of minimal coupling. If a finite inter-wire tunneling $t$ is turned on,  crossings between parabolas are lifted as shown by the solid lines in Fig.~\ref{fig:hall_levels} (a). The gap opened at the crossing of parabolas of neighboring wires is of order $t$, other crossings are lifted by higher-order tunnelings. If the chemical potential is tuned to an energy window with anticrossings, the coupled-wire system exhibits a bulk gap and chiral gapless edge states. We can thus identify the array of wires to be in a quantum Hall state. This is also illustrated by the striking similarity of Fig.~\ref{fig:hall_levels} (a) with Fig.~\ref{fig:hall_levels} (b) depicting the spectrum of the Hall effect in the limit of an isotropic two-dimensional electron gas.

The analysis of quantum Hall states in weakly coupled chains and other quasi-one-dimensional systems dates back to the 1980s. Already the seminal study by Thouless, Kohmoto, Nightingale, and den Nijs (TKNN) \cite{tknn1982} discusses the physics of quantum Hall states in anisotropic systems in the language of a two-dimensional lattice with unequal tunneling amplitudes along $x$ and $y$. Also network model for quantum Hall effects can, in their anisotropic versions, be understood as close relatives of arrays of coupled quantum wires \cite{cc1988,d1994}. An experimental motivation for the analysis of anisotropic quantum Hall states is provided by organic Bechgaard salts (TMTSF)$_2$X, where TMTSF stands for tetramethyselenafulvalene and X is a monovalent anion, as signatures of integer and fractional quantum Hall effects have been observed in those compounds \cite{cckbmneg1983,rcjmmb1984,cnychc1988,ckamjb1989,hbkcc1989,bkw1995,mz1997}. All of these studies underlines that isotropy is simply not relevant for the topological features of the quantum Hall effect.

\section{Coupled-wire constructions of fractional quantum Hall states}\label{sec:cw}
Since they require strong electron-electron interactions, fractional quantum Hall states cannot be described in a non-interacting framework. Coupled-wire constructions provide a particularly elegant way to treat interactions. Following a seminal study by Kane, Mukhopadhyay, and Lubensky \cite{kml2002}, this section exemplifies how coupled-wire constructions describe the so-called Laughlin states at filling factors $\nu=1/(2m+1)$ with integer $m$ \cite{laughlin_state}.

In a non-interacting picture, a quantum Hall system at filling $\nu=1/(2m+1)$ is gapless: only one in $2m+1$ states in the lowest Landau level is filled. The chemical potential is  located in the middle of the lowest Landau level, and the single-particle spectrum is does not have a gap. Experimentally, however, quantum Hall samples often exhibit a bulk gap and gapless, chiral edge states at fractional filling factors \cite{tsg1982}. The edge states support a fractional Hall conductance $\sigma_{yx} = \nu\,e^2/h=(2m+1)^{-1}\,e^2/h$ \cite{tsg1982}. Even more more intriguingly, bulk quasiparticles above the gap  carry a fractional charge $q= \nu\,e = (2m+1)^{-1}\,e$ \cite{laughlin_state,wen_book,gs1995,sgje1997,dp1997}. 

\begin{figure}
\centering
\includegraphics[height=3cm]{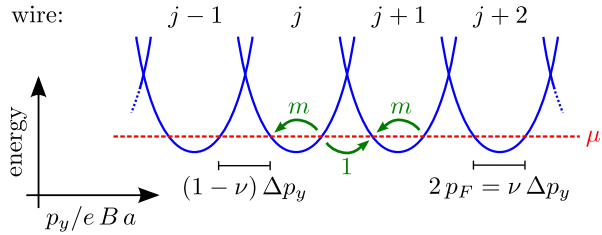}
  \caption{Dispersion of decoupled wires at a general filling factor $\nu<1$. At $\nu=1/(2m+1)$ with integer $m$, the correlated tunneling process depicted by the green arrows preserves the total momentum (the labels indicate the number of electrons tunneling along an arrow).}
  \label{fig:fqhe}
\end{figure}

Our starting point for the coupled-wire construction of Laughlin states is the array of wires introduced in Sec.~\ref{sec:anisotropy} and in Eq.~\eqref{eq:hamcwnonint}: a collection of wires extended along $y$ with inter-wire distance $a$. The wires are exposed to an out-of-plane magnetic field $\mathbf{B}=B\,\mathbf{e}_z$. As we shall see now, coupled-wire constructions technically correspond to a perturbative analysis in the inter-wire tunneling $t$. Let us thus shortly characterize the decoupled case $t=0$ in which the spectrum consists of parabolas shifted by the vector potential. Introducing $p_F = \sqrt{2\,m\,\mu}$, the Fermi momenta in wire $j$ are  
$p_{FRj} = p_F + e\,B\,j\,a$ for right-movers and $p_{FRj} = -p_F + e\,B\,j\,a$ for left-movers. In analogy to the integer quantum Hall case discussed in Sec.~\ref{sec:anisotropy}, we define the filling factor $\nu=1$ as the situation in which the chemical potential is at the energy of the crossings of dispersions of neighboring wires. For other chemical potentials, the filling factor is given by $\nu = 2\,p_F/\Delta p_y$, where $ \Delta p_y = e\,B\,j\,a$ is the momentum spacing between the dispersions of neighboring wires resulting from minimal coupling. The Fermi points of right-movers in wire $j$ and left-movers in wire $j+1$ thus have a momentum difference of $p_{FLj+1}-p_{FRj} =  (1-\nu)\,\Delta p_y$. 

\begin{figure}
\centering
\hspace*{1cm}\raisebox{2.75cm}{(a)}~\includegraphics[height=3cm]{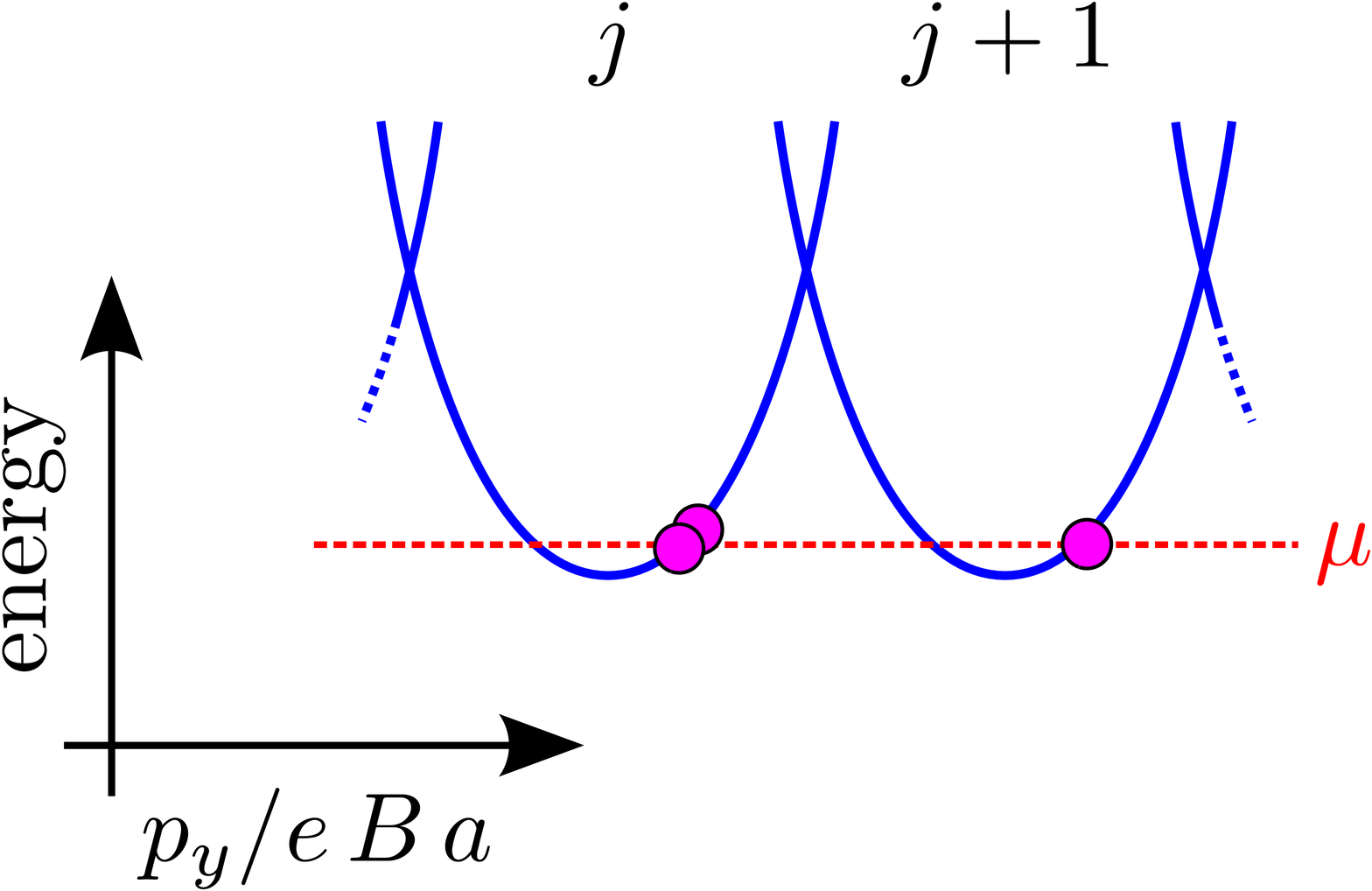}\hspace*{1.5cm}\raisebox{2.75cm}{(b)}~\includegraphics[height=3cm]{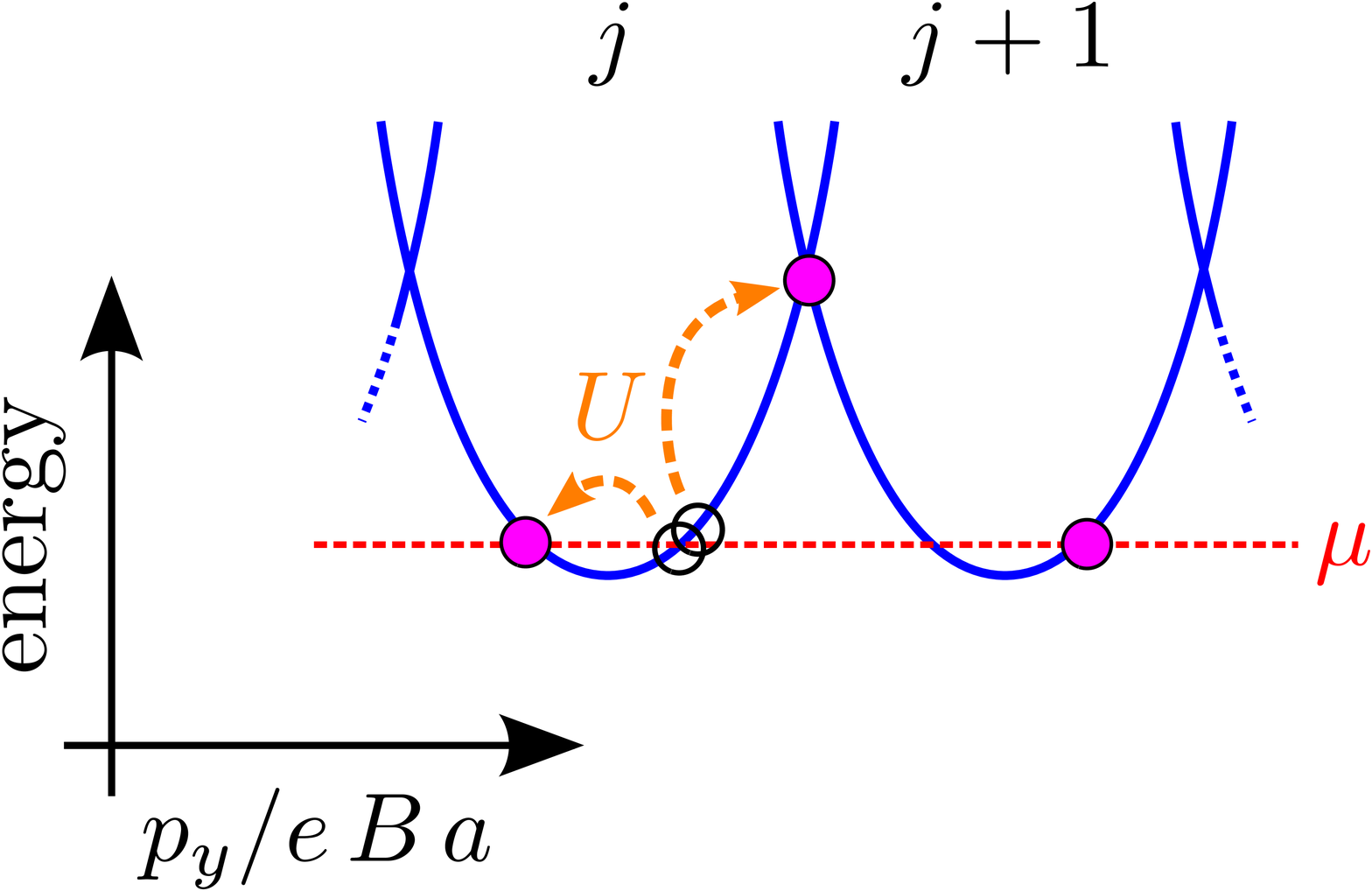}\\[0.5cm]
\hspace*{1cm}\raisebox{2.75cm}{(c)}~\includegraphics[height=3cm]{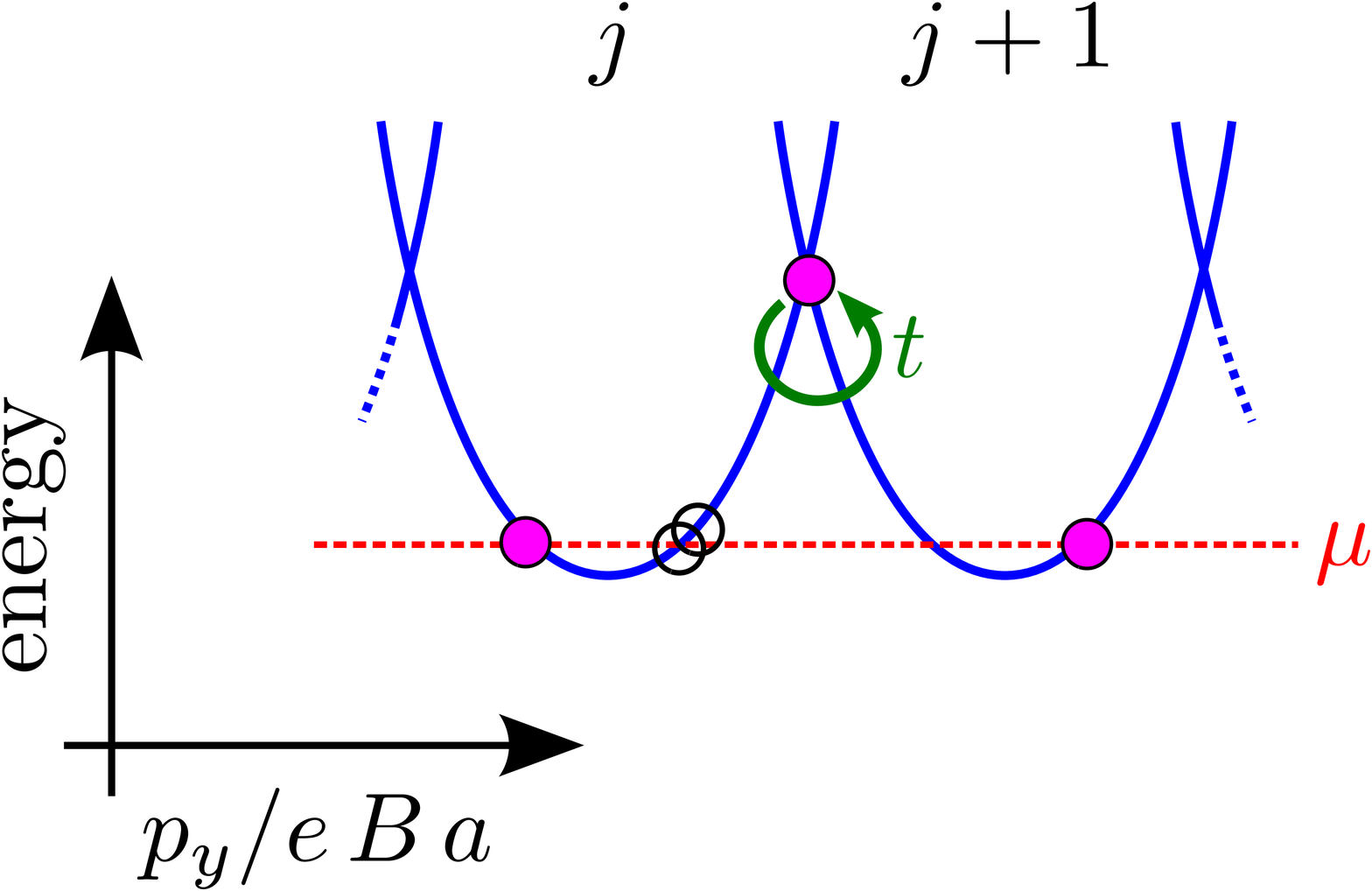}\hspace*{1.5cm}\raisebox{2.75cm}{(d)}~\includegraphics[height=3cm]{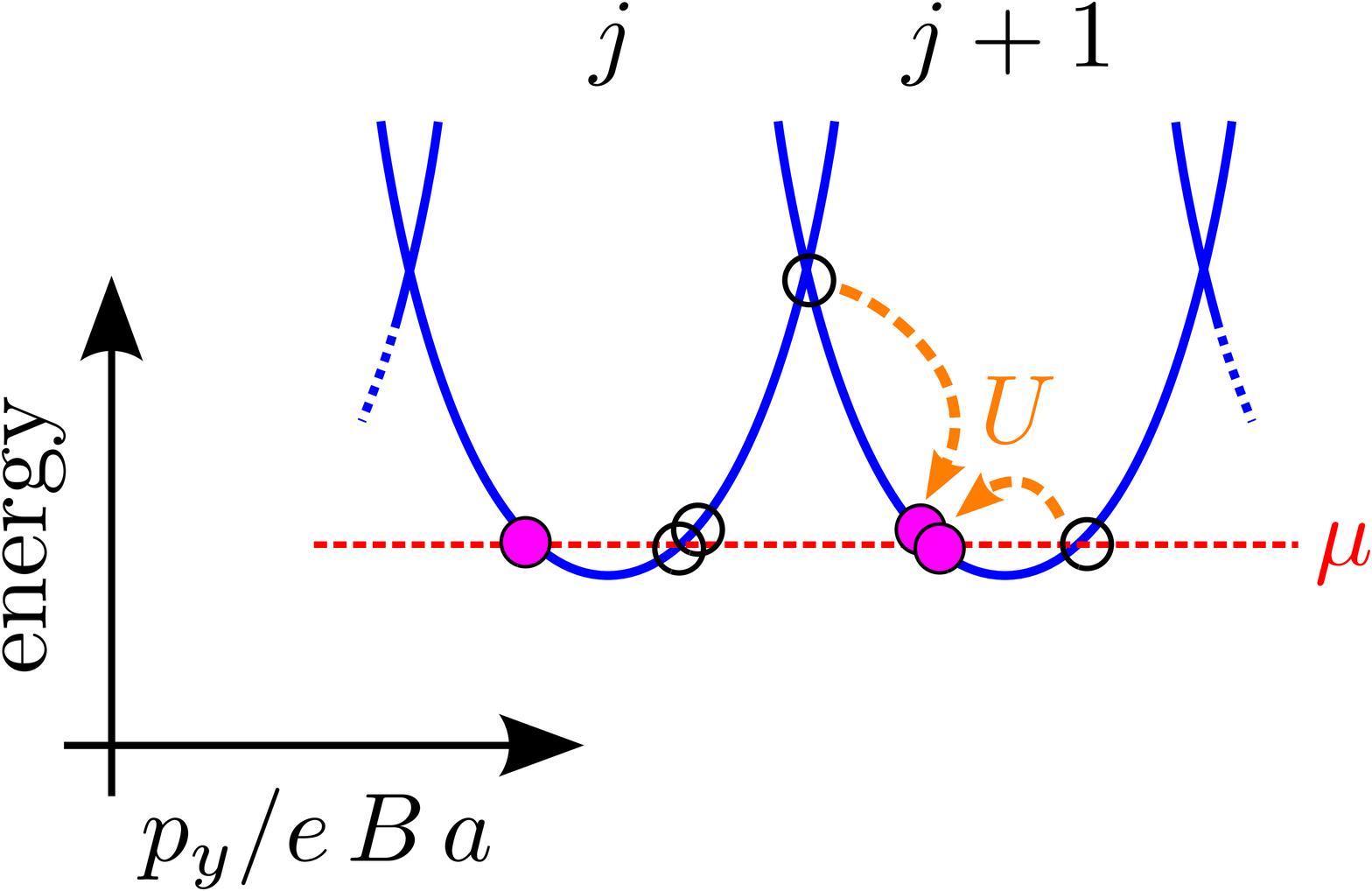}
  \caption{Generation of the correlated tunneling process that drives the system into a $\nu=1/3$-Laughlin state (corresponding to $m=1$ in Fig.~\ref{fig:fqhe}). Filled circles represent electrons, open circles indicate the location of the electrons at earlier steps. Panel (a) shows the initial situation. Panel (b) depicts the action of a momentum-conserving two-particle interaction $U$ in wire $j$ leaving the system in an intermediate virtual state, panel (c) shows the tunneling of one electron between wires $j$ and $j+1$, while a second interaction process in wire $j+1$ brings the system into the final state as shown in panel (d).}
  \label{fig:processes}
\end{figure}

For $\nu\neq1$, the anti-crossings opened by single-particle tunneling are not located at the Fermi level. Weak single-particle tunneling is then irrelevant for the low-energy physics. The combined presence of single-particle tunneling and electron-electron interactions, however, allows correlated tunneling events at the Fermi level. The process depicted in Fig.~\ref{fig:fqhe} is for example generated in $2m$-th order of an electron-electron interaction $U$. Its generation is detailed in Fig.~\ref{fig:processes} for $m=1$ ($\nu=1/3$). Such a process conserves the total momentum if the momentum transfer of the backscattering processes compensates the momentum transfer of inter-wire tunneling, and thus if $4\,m\,p_F = p_{FLj+1}-p_{FRj}$. This is precisely the case at filling factor $\nu = 1/(2m+1)$ \cite{kml2002}.

To study these correlated tunneling processes, we linearize the spectrum of each wire around its Fermi points and decompose the operator $c_{j}(y)$ annihilating an electron at position $y$ in wire $j$ into its right-moving ($R$) and left-moving ($L$) components as $c_{j}(y)\approx e^{-ip_{FLj}y}L_{j}(y)+e^{ip_{FRj}x}R_{j}(y)$. Following the conventions of Ref.~\cite{giamarchi_book}, we bosonize the chiral operators as $r_{j}(y) =(U_{r j}/\sqrt{2\pi\alpha})\,e^{-i\Phi_{r j}(y)}$, where $r=R,L \equiv +1,-1$, while $\alpha^{-1}$ is a large momentum cutoff, and $U_{r j}$ denotes a Klein factor.  The bosonized fields satisfy the commutator $[\Phi_{r j}(y),\Phi_{r' j'}(y')]=\delta_{rr'}\delta_{jj'}\,i\pi r\,\text{sgn}(y-y')$. In the following, we drop the Klein factors $U_{r j}$ to simplify the notation. They are not important for our discussion \cite{ds1998,kml2002}, and can be restored if needed.

The tunneling process depicted in Fig.~\ref{fig:fqhe} correspond to the fermionic Hamiltonian
\begin{align}
H_{j,j+1}^{(2m+1)}=\widetilde{g}_{2m+1}\,&\int dy\,\prod_{p=1}^m\left([\partial_y^pL_{j}^\dagger(y)]\,[\partial_y^{p-1}R_j^\pd(y)]\right)\,\prod_{q=1}^m\left([\partial_y^{q-1}L_{j+1}^\dagger(y)]\,[\partial_y^qR_{j+1}^\pd(y)]\right)\nonumber\\
&\times L_{j+1}^\dagger(y)\,R_{j}^\pd(y)+\rm{h.c.}.\label{eq:tun_ham}
\end{align}
The derivatives encode  small displacements of the operators: because Pauli principle entails $r_j(y)\,r_j(y)=0$,  a Taylor expansion for small displacements $\eta$ yields $r_j(y)\,r_j(y+\eta)\approx \eta\,r_j(x)\,\partial_y\,r_j(y)$. To derive the bosonized expression of $H_{j,j+1}^{(2m+1)}$, one can explicitly keep the small displacements in the fermionic creation and annihilation operators, then bosonize these operators, and finally obtain the leading contribution to the bosonized form of $H_{j,j+1}^{(2m+1)}$ by viewing the displacements as a point splitting. This yields

\begin{align}
&H_{j,j+1}^{(2m+1)}=g_{2m+1}\,\int dy \,\cos\Bigl([m+1]\,\Phi_{Rj}+m\,\Phi_{Lj}-m\,\Phi_{Rj+1}-[m+1]\,\Phi_{Lj+1}\Bigr).\label{eq:fqhe_sine_gordon0}
\end{align}
To simplify the notation in the remainder, we introduce new fields

\begin{align}
\widetilde{\Phi}_{Rj} = (1+m)\,\Phi_{Rj}-m\,\Phi_{Lj}\quad\text{and}
\quad\widetilde{\Phi}_{Lj} = (1+m)\,\Phi_{Lj}-m\,\Phi_{Rj}.\label{eq:new_fields}
\end{align}
The sine-Gordon terms can then be rewritten as 
\begin{align}
H_{j,j+1}^{(2m+1)}=g_{2m+1}\,\int dy\,\cos\left(\widetilde{\Phi}_{Rj} -\widetilde{\Phi}_{Lj+1} \right).\label{eq:fqhe_sine_gordon}
\end{align}
The Laughlin state at $\nu=1/(2m+1)$ corresponds to the strong-coupling phase of $\sum_jH_{j,j+1}^{(2m+1)}$, which is realized if all $H_{j,j+1}^{(2m+1)}$ flow to strong coupling in a renormalization group (RG) approach. The derivatives in the fermionic expression in Eq.~\eqref{eq:tun_ham} show that $H_{j,j+1}^{(2m+1)}$ is strongly irrelevant at the non-interacting fixed point. We thus not only need interactions to generate $H_{j,j+1}^{(2m+1)}$ in the first place, but also require particularly strong interactions to make it relevant.  Luckily, the microscopic interactions can always be fine-tuned such that this is the case.  To see this, we introduce new fields

\begin{align}
\widetilde{\phi}_{j+1/2} &= \frac{\widetilde{\Phi}_{Rj}-\widetilde{\Phi}_{Lj+1}}{2} \quad \text{and}\quad \widetilde{\theta}_{j+1/2} = \frac{-\widetilde{\Phi}_{Rj}-\widetilde{\Phi}_{Lj+1}}{2\,(2m+1)},
\end{align}
which obey the canonical commutator  $[\widetilde{\phi}_{j+1/2}(y),\widetilde{\theta}_{j'+1/2}(y')] = \delta_{jj'}\,(i\pi/2)\,\text{sgn}(y'-y)$. One can now always demand (fine-tune) the quadratic part of the Hamiltonian to be of the form
\begin{align}
H_{\rm quadr.}=\int dy\sum_j\left(\frac{u}{K}(\partial_y\widetilde{\phi}_{j+1/2})^2+u\,K\,(\partial_y\widetilde{\theta}_{j+1/2})^2\right)\label{eq:hamsll}
\end{align}
with $K<2$. This in turn renders the sine-Gordon term RG-relevant. 

Expressed in terms of an interacting fermionic Hamiltonian, the microscopic interactions that realize $H_{\rm quadr.}$ in Eq.~\eqref{eq:hamsll} are of an admittedly special type. Let us nevertheless assume that this is the case. We furthermore focus on an array on $n$ wires, such that the full Hamiltonian is given by
\begin{align}
H=H_{\rm quadr.}+\sum_{j=1}^{n-1} H_{j,j+1}^{(2m+1)}.
\end{align}
When all $H_{j,j+1}^{(2m+1)}$ flow to strong coupling, all modes in the bulk of the system are gapped. The chiral modes $\widetilde{\Phi}_{L1}$ and $\widetilde{\Phi}_{Rn}$ located at the edges of the system, however, do not appear in any of the sine-Gordon terms. Being unrestrained, these two modes remain gapless. 

We have thus constructed a bulk gapped state at filling factor $\nu=1/(2m+1)$ with one gapless chiral mode per edge that fundamentally requires the presence of strong electron-electron interactions. All of these properties match the Laughlin states \cite{laughlin_state,wen_book}. To unambiguously identify the strong-coupling phase of $\sum_{j=1}^{n-1} H_{j,j+1}^{(2m+1)}$ as a $\nu=1/(2m+1)$-Laughlin state, we should also recover the characteristic fractionalized bulk quasiparticles of charge $e/(2m+1)$. These quasiparticles are the minimal excitations of the Laughlin states above the bulk gap. In a coupled-wire language, a minimal bulk excitation corresponds to a kink in one of the sine-Gordon terms: in its strong-coupling phase, the ground state of
$$
H_{j_0,j_0+1}^{(2m+1)}=g_{2m+1}\,\int dy\,\cos\left(\widetilde{\Phi}_{Rj_0} -\widetilde{\Phi}_{Lj_0+1} \right)
$$
has $\widetilde{\Phi}_{Rj_0} -\widetilde{\Phi}_{Lj_0+1}$ pinned to one of the minima of the cosine. A minimal excitation at $y=y_0$ arises if the fields are not strictly pinned, but interpolate between two neighboring minima as
\begin{align}
\widetilde{\Phi}_{Rj_0}(y) -\widetilde{\Phi}_{Lj_0+1}(y)=\begin{cases}\widetilde{\Phi}_0&\text{for }y<y_0,\\\widetilde{\Phi}_0+2\pi&\text{for }y>y_0,\end{cases}
\end{align}
where $\widetilde{\Phi}_0$ corresponds to one of the minima of the sine-Gordon term.
To determine the charge associated with such a kink, we study the total charge $Q$ in the system. Using Eq.~\eqref{eq:new_fields}, we can express $Q$ as the integral of the one-dimensional charge densities,
\begin{align}
Q &=  -\frac{e}{2\pi}\sum_{j=1}^n\,\int dy\,\partial_y\left(\Phi_{Rj}(y)-\Phi_{Lj}(y)\right)\nonumber\\
&= -\frac{e}{2\pi}\sum_{j=1}^n\,\int dy\,\partial_y\frac{\widetilde{\Phi}_{Rj}(y)-\widetilde{\Phi}_{Lj+1}(y)}{2m+1}+\text{boundary terms}.
\end{align}
The charge associated with a kink in $\widetilde{\Phi}_{Rj_0}(y) -\widetilde{\Phi}_{Lj_0+1}(y)$ is thus
\begin{align}
Q_{\rm kink}&= -\frac{e}{2\pi}\int dy\,\partial_y\frac{\widetilde{\Phi}_{Rj_0}(y)-\widetilde{\Phi}_{Lj_0+1}(y)}{2m+1} = -\frac{e}{2m+1}.
\end{align}
Combined with the correct filling factor, the bulk gap, and the chiral edge states, the fractionally charged quasiparticles show that the strong-coupling phase of the coupled-wire construction is a Laughlin state at filling $\nu=1/(2m+1)$.

While this is great news, one important question remains: can we find a physical system in which $H_{j,j+1}^{(2m+1)}$ is important at all? As  discussed above, forward scattering interactions need to be of a rather specific type to realize a Laughlin state. Even worse, the strong-coupling phase of $H_{j,j+1}^{(2m+1)}$ is only reached if there is no competing operator that flows to strong coupling faster that the correlated tunnelings, and for instance drives the system towards superconducting instabilities or into a density wave state \cite{kml2002,skl2016}. If we did not know that fractional quantum Hall states exist, we would probably dismiss $H_{j,j+1}^{(2m+1)}$  as physically completely irrelevant.  

Despite being in the same topological state, however, experimental fractional quantum Hall systems typically do not realize the extremely anisotropic version of a Laughlin state associated with $H_{j,j+1}^{(2m+1)}$. One should rather view coupled-wire constructions as the analogue of a Luther-Emery point \cite{le1974,giamarchi_book} in the phase-space of fractional quantum Hall states: a special, very anisotropic point at which the system can be solved more or less exactly. Coupled-wire constructions assert that this point is not a singular one, but part of an extended phase including much more isotropic systems. Being topological, it is indeed natural to expect fractional quantum Hall state to be robust to changes in the Hamiltonian. Concomitantly, we saw in Sec.~\ref{sec:anisotropy} that the integer quantum Hall effect  survives the tuning of anisotropy from a two-dimensional electron gas to weakly coupled quantum wires. Similarly, the experimentally observed Laughlin states are believed to be smoothly connected to the special points at which coupled-wire Hamiltonians describe them.

\section{Scope of coupled-wire constructions}\label{sec:review}
Building on the insightful initial work by Kane, Mukhopadhyay, and Lubensky \cite{kml2002}, coupled-wire constructions have been extended to many other topological states. An important generalization by Teo and Kane in Ref.~\cite{tk2014} showed  that coupled-wire constructions provide a simple theoretical framework for non-Abelian quantum Hall states. This work also discusses how the nontrivial braiding properties of bulk quasiparticles are reflected in coupled-wire constructions. Since then, coupled-wire constructions have been used to reproduce an impressive number of interacting topological states of matter. Symmetry-protected topological phases with additional symmetries were studied in Ref.~\cite{la2012}. The topological degeneracy of fractional quantum Hall states on a torus was detailed in Ref.~\cite{sosh2015}. The Chern-Simons action for the electromagnetic field, which emerges as the low-energy description of quantum Hall states when the electrons are integrated out, has been derived in Ref.~\cite{shcgg2015}. Flux attachment has been discussed in Ref.~\cite{mam2017}, and a detailed overview of its implementation for various fractional quantum Hall states has been given in Ref.~\cite{ff2019}. An alternative vision of the connection of the coupled-wire approach and low-energy field theory of fractional quantum Hall states was given in Ref.~\cite{ith2019}, which also discusses how quantum Hall wave functions can be distilled from the coupled-wire approach. The tenfold way classification of non-interacting topological states has been reproduced in Ref.~\cite{ncmt2014}, where also strongly interacting phases with short-range entanglement beyond the tenfold-way classification and long-range entangled topological phases have been constructed.

Building on these insights, coupled-wire constructions  have been employed for a comprehensive analytic description of an ever-growing list of fractional quantum Hall states and related systems since they provide rather straightforward access to fascinating physics \cite{kl2013,jt2014,so2014,ms2014,kl2014,vb2014,mskl2014,ktl2015,bmf2015,chm2015,ksl2016,mrf2016,fl2017,hg2017,ks2018,tkl2018,ssct2018,bzpg2019,bfg2019,fgh2019}. Coupled-wire constructions are, however, not restricted to fermionic and bosonic topological  states, but may also describe topological states in spin models, the so-called spin liquids. As for quantum Hall states, coupled-wire constructions allow for a description of a broad range of these states \cite{ssl2002,nt2003,mngt2015,gps2015,hcgncm2016,fhbp2016,pc2016,hkt2017,hcfcm2017,lt2017,cmct2017,pb2018,cmci2019}. Also paired states of matter and topological superconductivity, which has important applications in quantum computation, have been studied in a coupled-wire language \cite{gang11,v2014, sbo2014,szt2016,shboo2017,ksh2017,prgt2018,llk2019}.

Beyond the reproduction of known results by a powerful alternative method, coupled-wire constructions also allow to explore new physics. An example are one-dimensional analogues of quantum Hall states that can be understood as coupled-wire constructions in the limit of small arrays of quantum wires. The minimal system that allows for a sine-Gordon term of the form of Eq.~\eqref{eq:fqhe_sine_gordon0} consists of two spinless quantum wires, or alternatively one spinful wire. In its strong-coupling phase, such a sine-Gordon term gives rise to a one-dimensional analogue of a fractional quantum Hall state dubbed a fractional helical Luttinger liquid \cite{oss2014}. Among other intriguing properties, this  purely one-dimensional state is characterised by a fractional conductance \cite{oss2014,mfsl2014}. One-dimensional analogues of fractional quantum Hall states have by now turned into an active field of research on their own \cite{gh2014,cns2015,cgbs2015,cs2015,btrmf2015,zwz2015,pl2015,ccrbdfsm2017,belsbz2017,pprml2017,alk2018,hrb2018,fbg2018,so2019,ssss19}. Particularly exciting are heterostructures of those states with superconductors since they can host so-called parafermionic zero modes at domain walls, and may also exhibit an $8\pi$-periodic Josephson effect \cite{lbrs2012,cas2013,bjq2013,kyl2014,kl2014pf,bq2014,zk2014,kl2014pf2,mgcgs2015,otms2015,kl2015pf,cb2016,pmts2016,tfdt2017,esto2017,pmts2017,acns2017,vbo2017,sh2017pf,vv2017,cmss2018,fztt2019}.

An exciting future perspective for coupled-wire constructions is their application to interacting three-dimensional systems. These are particularly hard to tackle by other approaches: three-dimensional systems are numerically costly, and the analytic description of strongly interacting three-dimensional states is notoriously hard. Coupled-wire constructions offer a powerful alternative approach, and have already been used to discuss several strongly interacting topological states both on the surface and in the bulk of three- and higher-dimensional systems \cite{amorf2009,bs2009,vs2013,v2013,jq2014,orfm2014,so2015,m2015,if2015,mea2015,mgsb2016,szt2016,mam2016_2,meas2016,tlk2016,incm2016,skl2016,incm2017,lsl2017,y2017,vlk2017,hf2017,c2018,ht2018,prgt2018,ssm2018,ff19}. Future work should explore coupled-wire constructions in three spatial dimensions in greater detail. The existing literature lays out two particularly promising directions. On the one hand, one can study bulk gapped phases in three-dimensional coupled-wire systems built from globally commuting sets of sine-Gordon terms. Here, true three-dimensionality can be achieved if the sine-Gordon terms not only couple pairs of neighboring wires, but for example form star and plaquette operators in the two-dimensional lattice that the wires form perpendicular to their extended direction. On the other hand, fractionalized three-dimensional phases can also emerge from coupled layers in which three-dimensional coherence arises as a result of anyon condensation. In this case, coupled-wire constructions serve to describe the strongly-interacting physics in the underlying coupled layers.

\section{Coupled-wire constructions beyond two-dimensional quantum Hall states}\label{sec:general}
We conclude by demonstrating the versatility of coupled-wire constructions with two concrete examples beyond quantum Hall states.

\subsection{Coupled-wire description of chiral spin liquids}\label{subsec:outlook}
Unlike their name suggests, coupled-wire constructions are by no means restricted to electronic quantum wires. They describe higher-dimensional topological states arising from suitable couplings in any array one-dimensional subsystems. When the subsystems are spin chains, the resulting topological state is a spin liquid. For the so-called chiral spin liquid, a coupled-wire construction has been introduced in Ref.~\cite{mngt2015} (a similar idea has been pursued in \cite{gps2015}). 

Chiral spin liquids can be understood as a fractionally quantized Hall liquid for bosonic spin flip operators acting on a spin-polarized reference
state  \cite{kalmeyer-87prl2095,kalmeyer-89prb11879,wen-89prb11413,laughlin-90prb664}. The fractional charge associated with quasiparticles in electronic fractional quantum Hall states is replaced by the existence of spinons with spin $S=1/2$ in a system whose microscopic excitations are spin flips carrying a spin  $S=1$. To describe chiral spin liquids in a coupled-wire language, Ref.~\cite{mngt2015} started from an array of spinful quantum wires. Bosonization then gives rise to spin and charge modes, whose dynamics decouple due to spin-charge separation \cite{giamarchi_book}. When the charge sectors of each wire enter a Mott gap, the low-energy dynamics is determined by the spin sector only. The corresponding Hamiltonian is related to the one of a Jordan-Wigner transformed Heisenberg spin chain \cite{giamarchi_book}: projected to the spin sector, the array of Mott-gapped wires corresponds to a collection of single-component Luttinger liquids. The chiral spin liquid arises as the strong-coupling phase of specific inter-chain spin-flip terms that take the form of effective hoppings between the Luttinger liquids describing the spin sectors of neighboring Mott-gapped wires. Spinons with spin $S=1/2$ are then described by kinks in the arguments of the corresponding sine-Gordon terms.

\subsection{Fractional chiral metals: coupled-Weyl node constructions in higher dimensions}\label{subsec:outlook2}
The band structures of three-dimensional systems generically exhibit so-called Weyl nodes: gapless points in momentum space at which pairs of non-degenerate bands touch \cite{h1937,volovik_book,wtvs2011}. For recent reviews of Weyl semimetal physics, see for example Refs.~\cite{tv2013,yf2017,b2018,amv2018}. If the Weyl nodes are located at the Fermi energy, Weyl semimetals can be described by an expansion of the Hamiltonian around these special points. To linear order in the three-dimensional momentum $\mathbf{p}$ measured relative to the Weyl node, the low-energy expansion reads
\begin{align}
H_{\rm Weyl}^\chi &= \sum_{\mathbf{p}}\left(c_{\mathbf{p}\uparrow}^\dagger,c_{\mathbf{p}\downarrow}^\dagger\right)\,\mathcal{H}_{\mathbf{p}}^\chi\,\begin{pmatrix}c_{\mathbf{p}\uparrow}^\pd\\c_{\mathbf{p}\downarrow}^\pd\end{pmatrix}\quad\text{with}\quad \mathcal{H}_{\mathbf{p}}^\chi = \chi\,v_F\,\mathbf{p}\cdot\boldsymbol{\sigma},\label{eq:weyl_ll_ham}
\end{align}
where $c_{\mathbf{p}\sigma}^\dagger$ creates an electron with spin $\sigma=\uparrow,\downarrow$, while $v_F$ is the Fermi velocity, $\boldsymbol{\sigma}$ denotes the vector of Pauli matrices, and $\chi=\pm1$ is the chirality of the Weyl node. The chirality corresponds to whether momentum $\mathbf{p}$ and spin $\boldsymbol{\sigma}$ are aligned or anti-aligned in the state of lowest energy. Weyl semimetals are especially interesting because of their topological character: it can be shown that Weyl nodes are monopoles of Berry curvature, and that the sign of the monopole's Berry-charge equals the chirality of the node. 

\begin{figure}
  \centering
  \centering
  \includegraphics[width=0.25\columnwidth]{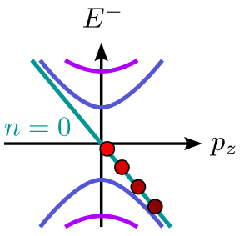}\quad\quad  \includegraphics[width=0.25\columnwidth]{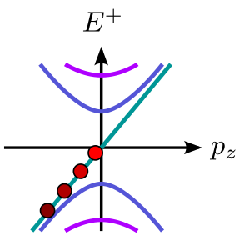}\\\vspace*{0.35cm}\scalebox{2}{$\downarrow$}\hspace*{0.4cm}\text{electric field }$\mathbf{E}\parallel \mathbf{B}$\hspace*{0.4cm}\scalebox{2}{$\downarrow$}\hspace*{0.5cm}~\\\vspace*{0.5cm}
 \includegraphics[width=0.25\columnwidth]{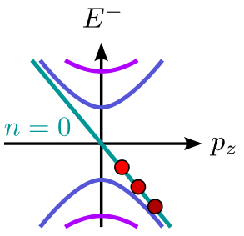}\quad\quad  \includegraphics[width=0.25\columnwidth]{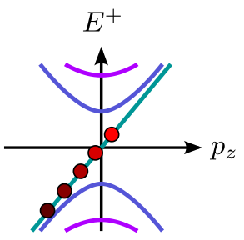}
  \caption{Landau levels and the chiral anomaly in Weyl semimetals. Top row: dispersion of the Landau levels as a function of momentum $p_z$, and the occupation of the $n=0$ Landau level with electrons (filled circles) in equilibrium. Bottom row: an electric field $\mathbf{E}$ parallel to the $z$-axis increases the $z$-momentum of the electrons, thereby changing the number of electrons at a given Weyl node while keeping the total electron number in a pair of Weyl nodes constant.
}
  \label{fig:weyl_b_4d}
\end{figure}

The topological character of Weyl nodes leads to a many fascinating consequences, including maybe most prominently the so-called chiral anomaly in applied electric and magnetic fields. In a semiclassical picture, a magnetic field along the $z$-axis forces electrons onto cyclotron orbits in the $(x,y)$-plane, but free motion is still possible parallel to the field. In a full quantum-mechanical calculation, one finds that Weyl semimetals in a magnetic field develop Landau levels similar to the two-dimensional electron gases discussed in Sec.~\ref{sec:anisotropy}. The main difference arising from the existence of a third dimension is that the Landau levels now disperse with the momentum parallel to the field. For $\mathbf{B}=B\,\mathbf{e}_z$, the dispersion $E_n^{\chi}$ of the Landau level with index $n$ is given by
\begin{align}
E_n^{\chi}(p_y,p_z) = \begin{cases}\text{sgn}(n)\,\sqrt{v_F^2\,p_z^2+2\,|e\,B|\,n}&\text{ for }n\neq0,\\
\chi\,\text{sgn}(e\,B)\,v_F\,p_z&\text{ for }n=0.\end{cases}
\end{align}
The Landau level spectrum of Weyl nodes is shown in Fig.~\ref{fig:weyl_b_4d}. Up to the macroscopic Landau level degeneracy $N_{LL} = L_x\,L_y\,e\,B/(2\pi)$ due to $p_y$ (where $L_{x(y)}$ is the system size in $x$($y$)-direction), the zeroth Landau level is similar to a one-dimensional mode of chirality $\chi\,\text{sgn}(eB)$. In the low-energy approximation, the zeroth Landau level continuous to disperse linearly up to negative infinite energy, and all states below zero energy are occupied in equilibrium. The Weyl Hamiltonian hence comes with an infinite reservoir of occupied electronic states below the Fermi level. \\~\\
Consider now applying an electric field $\mathbf{E} = E\,\mathbf{e}_z$ parallel to the magnetic field. In the resulting non-equilibrium situation, the electric field changes the  momenta of electrons as $\dot{p}_z =e\,E$, see Fig.~\ref{fig:weyl_b_4d}. Because there is an infinite reservoir of electrons at negative energies,  the net result is to change the charge density of fermions with chirality $\chi$ as

\begin{align}
\dot{\rho}_{\rm charge} &= \underbrace{e\,\frac{1}{L_x\,L_y\,L_z}}_{=\text{charge density}}\,\underbrace{\chi\,\text{sgn}(e\,B)}_{=\text{effective 1D chirality}}\,\underbrace{\frac{L_x\,L_y\,|e\,B|}{2\pi}}_{=\text{degeneracy}}\,\underbrace{\frac{L_z}{2\pi}}_{=1/\Delta p_z}\,\underbrace{eE}_{=d p_z/d t}=\chi \,\frac{e^3}{4\pi^2}\,\mathbf{E}\cdot\mathbf{B},\label{eq:chiral_anomaly}
\end{align}
To reconcile the chiral anomaly with global electron number conservation, which we physically have to impose in any solid, Weyl semimetals always feature pairs of Weyl nodes of opposite chirality with $\sum_\chi \dot{\rho}_{\rm charge}^\chi = 0$ \cite{nn1981a,nn1981b,nn1981c,nn1983,mb2019}. The chiral anomaly can then be interpreted as a pumping of electrons from one Weyl node to another.

\begin{figure}
  \centering
    \includegraphics[height=3cm]{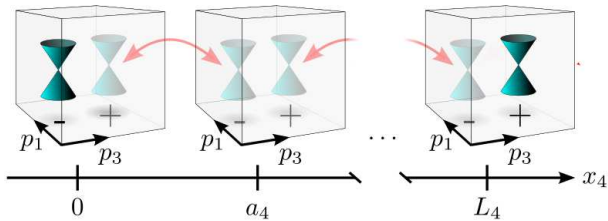}
  \caption{Construction of an integer quantum Hall effect in four spatial dimensions from coupled Weyl semimetals. A tunnel coupling between Weyl nodes of opposite chirality $\chi=\pm1$ in neighboring Weyl semimetals, depicted by red arrows, induces a gap for all bulk nodes. In a slab of finite extent $0\leq x_4\leq L_4$, single gapless nodes remain at the three-dimensional surfaces at $x_4=0$ and  $x_4=L_4$. Gapped nodes are indicated by a fading, the signs below the nodes denote their chiralities. Adapted from Ref.~\cite{mgsb2016}.
  }
  \label{fig:4dweyl_coupled}
\end{figure}

As illustrated with the example of integer and fractional quantum Hall states in Sec.~\ref{sec:cw}, bulk-gapped non-interacting topological states often have strongly-interacting analogues exhibiting fractionalization. Given that Weyl nodes are topological but gapless, it is an exciting question if strong electron-electron interactions can stabilize fractionalized cousins of Weyl nodes. Those would be a gapless collection of three-dimensional states with definite chirality carrying a fractional electric charge. The fractional charge in turn implies a fractionalized response to electromagnetic fields, similar to the fractionalized Hall response $\sigma_{yx}=\nu\,e^2/h$ of a Laughlin state. In the context of Weyl nodes, this translates to a fractionalized variant of the chiral anomaly in Eq.~\eqref{eq:chiral_anomaly}. Ref.~\cite{mgsb2016} approached this problem by generalizing coupled-wire constructions to four spatial dimensions (4+1D). Like chiral modes in one dimension, two Weyl nodes of opposite chiralities gap out when being coupled. Coupled-wire constructions can thus be generalized to coupled-Weyl node constructions in 4+1D as shown in Fig.~\ref{fig:4dweyl_coupled}: the four spatial dimensions are viewed as three-dimensional subsystems stacked along a fourth direction $x_4$. If each of the three-dimensional subsystems contains a Weyl semimetal with two Weyl nodes of opposite chirality, a suitable coupling between the right-handed Weyl node at $x_4$ and the left-handed node at $x_4+a_4$ produces a state with a full bulk gap and one gapless Weyl node per edge ($a_4$ is the lattice spacing along the fourth direction). The combination of a 4+1D bulk gap and gapless three-dimensional edge states of definite chirality finally identify this state as a 4+1D integer quantum Hall state \cite{zh2001,kn2002,fp2000}.

To construct a fractional quantum Hall state in 4+1D, Ref.~\cite{mgsb2016} generalized coupled-wire constructions from two to four spatial dimensions by replacing the chiral modes of quantum wires with the chiral zeroth Landau levels of Weyl semimetals. As a starting point, consider subjecting the three-dimensional Weyl semimetals at each $x_4$ in a 4+1-dimensional stack of Weyl semimetals  to a magnetic field. In the limit of large fields, the low-energy physics is well-approximated by considering only the chiral zeroth Landau levels. 
  \begin{figure}
 \centering
  \includegraphics[height=3cm]{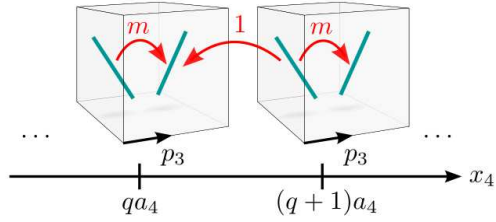}
  \caption{Correlated tunnelings between neighboring Weyl semimetals leading to fractional quantum Hall states. While an electron hops from the left-handed Weyl-node at $x_4=(q+1)\,a_4$ to the right-handed node at $q\,a_4$ (with $q\in\mathds{Z}$), $m$ electrons are scattered from the left-handed node to the right-handed node in both Weyl semimetals connected by the tunnelling. This correlated process is indicated by the arrows, whose labels indicate the number of electrons transported along the respective arrow. Taken from Ref.~\cite{mgsb2016}.
  }
  \label{fig:weyl_B_scattering}
\end{figure}
Next, consider correlated tunnelings between these quasi-one-dimensional modes similar to the ones stabilizing Laughlin states in two dimensions as illustrated in Fig.~\ref{fig:weyl_B_scattering}. Provided that there is no scattering between states with different $y$-momentum $p_y$ (the quantum number labelling the different degenerate states within a Landau level), the 4+1D stack of Weyl semimetal Landau levels can be mapped to $N_{LL}$ copies of coupled-wire constructions for quantum Hall states. One can then follow along the lines of two-dimensional coupled-wire constructions: each chiral Landau level is represented by $N_{LL}$ Luttinger liquids, and the correlated tunnelings shown in Fig.~\ref{fig:weyl_B_scattering} are described by sine-Gordon terms for these Luttinger liquids. The strong-coupling phase of these correlated tunnelings has a full bulk gap, and each (three-dimensional!) edge hosts a family of gapless modes of definite chirality (the family corresponds to the Landau-level-degenerate modes). The combination of a 4+1D bulk gap and gapless three-dimensional edge states of definite chirality identifies the strong-coupling phase of the correlated tunnelings as a 4+1D quantum Hall state. 

The fractionalized nature of this state is heralded by its response to applied electromagnetic fields. Generalizing the approach of Ref.~\cite{shcgg2015} to 4+1D, Ref.~\cite{mgsb2016}  considered an infinite 4+1D quantum Hall system and integrated out the gapped fermionic modes in the strong coupling-phase of the sine-Gordon terms. This yields an effective action describing the system's response to applied electromagnetic fields. For the 4+1D fractional quantum Hall state generated by the correlated tunnelings shown in Fig.~\ref{fig:weyl_B_scattering}, Ref.~\cite{mgsb2016} found the response to electromagnetic fields to be governed by the 4+1D Chern-Simons action
\begin{align}
\label{eq:CS4+1}
\mathcal{S}_\text{CS}^{(4+1)}[A_{\mu}]&=\frac{-e^3}{6(2\pi)^2(2m+1)} \int d^5 x \,\epsilon^{\mu\nu\rho\sigma\eta}\,A_\mu\partial_\nu A_\rho \partial_\sigma A_\eta,
\end{align}
where $A_\alpha$ is the $\alpha$-component of the vector potential 5-vector in (4+1)D. The current in the fourth direction flowing in response to applied electromagnetic fields is then given by $j^{4} ={\delta\mathcal{S}_\text{CS}^{(4+1)}}/{\delta A_4}$. If the system has a finite size  in the fourth direction, this current eventually hits the edge of the system and over time results in charge accumulating there. Using the definition of the electromagnetic fields as derivatives of the 5-vector potential $A_\mu$, this charge accumulation is given by
\begin{align}
\dot{\rho}_{\text{edge  charge}}&= \pm\frac{e^3}{4\pi^2(2m+1)}\,\mathbf{E}\cdot\mathbf{B}.
\end{align}
Because the accumulated charge must occupy low-energy states at the boundary, it has to end up in the chiral, gapless surface states. This means that the chiral, gapless surface states of a 4+1D fractional quantum Hall states react to applied electromagnetic fields by a fractionalized version of the chiral anomaly. The surface states can thus be identified as fractional analogues of single Weyl nodes; they have been dubbed fractional chiral metals.

While our physical world is only three-dimensional, higher-dimensional systems can be emulated in quantum simulators such as cold atomic gases with so-called synthetic dimensions \cite{bcll12,gbz16,op19}. The role of additional dimensions are then played by internal degrees of freedom. Different states $|i\rangle$ of the internal degrees of freedom are identified with lattice sites $i$ along the synthetic dimension, and internal transitions $|i\rangle\to|j\rangle$ play the role of hoppings in the synthetic direction. This concept has already been applied to the 4+1D integer quantum Hall effect \cite{pzocg2015,lspzb18}. Since cold atomic setups in principle also allow to control interactions \cite{bdz2008}, one can hope that cold-atom implementations of synthetic dimensions can in the future be developed to also host fractionalized higher-dimensional states of matter. Coupled-Weyl node constructions will then provide a comparably simple analytical description of such phases.

Still, the study of strongly interacting states in higher dimensions, for which fractional quantum Hall states in 4+1D are a paradigmatic example, is not only important to enlarge our understanding of topologically ordered states in general, but also allows us to learn more about strongly interacting three-dimensional states of matter. To illustrate this point, we recall that three-dimensional Weyl semimetals can be understood as slabs of four-dimensional integer quantum Hall states in which surface states, the Weyl nodes, are separated in momentum, but not in space. Similarly, a three-dimensional slab of a 4+1D fractional quantum Hall state can realize a fractional Weyl semimetal in 3+1D. The main technical benefit in thinking about 4+1D  fractional quantum Hall states is the fact that these exhibit a bulk gap, which in turn allows to relatively easily integrate out the electrons to obtain the response to electromagnetic fields. And since the topological response of a 4+1D bulk must be compensated by the topological response of its 3+1D edge, this means that one can relatively easily identify the response of gapless three-dimensional topological systems exhibiting fractionalization (we for example found that the current flowing in the bulk of the 4+1D fractional quantum Hall states corresponds to a charge accumulation at the edge). Given the success of coupled-wire constructions in two spatial dimensions, one can thus hope that coupled-Weyl node constructions can be developed into a similarly universal description of interacting topological states in 4+1D, and their descendants in 3+1D.

\section{Conclusions}\label{sec:conclusions}
Coupled-wire constructions use a Luttinger liquid picture and bosonization for the description of strongly interacting topological states in two and higher dimensions. As maybe most prominently heralded by the way gapless edge states emerge in them, coupled-wire constructions can be viewed as two-dimensional relatives of the one-dimensional Affleck-Kennedy-Lieb-Tasaki (AKLT) \cite{aklt1987} and Kitaev chains \cite{Kitaev_2001}. All of these constructions build topological phases by cleverly splitting and regrouping the original degrees of freedom. 

Coupled-wire constructions are able to analytically describe a large variety of higher-dimensional topological states at the expense of using a very anisotropic limiting case. This modelling technique is  particularly helpful for the analytical exploration of possible topological phases and their universal topological properties. Moving forward,  higher-dimensional variants of coupled-wire constructions are a particularly promising tool for the  study of topological systems in three (and higher) dimensions, a most exciting frontier in topological solid state physics.

\begin{acknowledgement}
We acknowledge financial support from the DFG via the Emmy Noether Programme ME 4844/1-1, SFB 1143 (project-id 247310070), and the W\"urzburg-Dresden Cluster of Excellence on Complexity and Topology in Quantum Matter - ct.qmat (EXC 2147, project-id 39085490).
\end{acknowledgement}

\bibliographystyle{epj}

\end{document}